\documentclass[lettersize,journal]{IEEEtran}
\usepackage{amsmath,amsfonts}
\usepackage{algorithmic}
\usepackage{algorithm}
\usepackage{array}
\usepackage{xcolor}
\usepackage[caption=false,font=normalsize,labelfont=sf,textfont=sf]{subfig}
\usepackage{textcomp}
\usepackage{bm}
\usepackage{stfloats}
\usepackage{url}
\usepackage{verbatim}
\usepackage{graphicx}
\usepackage{cite}

\begin{document}

\title{Predictive Control of EV Overnight Charging with Multi-Session Flexibility}

\author{
    Felix Wieberneit,
    \thanks{F. Wieberneit, H. Hamedmoghadam and R. Shorten are with the School of Design Engineering, Imperial College London.}
    \and
    Emanuele Crisostomi,
    \thanks{E. Crisostomi is with the Department of Energy, Systems, Territory and Constructions Engineering, University of Pisa.}
    \and
    Anthony Quinn,
    \thanks{A. Quinn is with the School of Design Engineering, Imperial College London, and with the Department of Electronic and Electrical Engineering, Trinity College Dublin.}
    \and
    Homayoun Hamedmoghadam,
    \and
    Robert Shorten
}

        % <-this % stops a space
%\markboth{}%

\IEEEpubid{This work has been submitted to the IEEE Transactions on Intelligent Transportation Systems.}

\maketitle

\begin{abstract}
The majority of electric vehicles (EVs) are charged domestically overnight, where the precise timing of power allocation is not important to the user, thus representing a source of flexibility that can be leveraged by charging control algorithms. In this paper, we relax the common assumption, that EVs require full charge every morning, enabling additional flexibility to defer charging of surplus energy to subsequent nights, which can enhance the performance of controlled charging.
In particular, we consider a simple domestic smart plug, scheduling power delivery with the objective to minimize CO$_2$ emissions over prediction horizons of multiple sessions---up to seven days ahead--- utilising model predictive control (MPC). 
Based on carbon intensity data from the UK National Grid, we demonstrate significant potential for emission reductions with multi-session planning of 40 to 46\% compared to uncontrolled charging and 19 to 26\% compared to single-session planning. Furthermore, we assess, how the driving and charging behaviour of EV users affects the available flexibility and consequentially the potential for emission reductions. Finally, using grid carbon intensity data from 14 different UK regions, we report significant variations in absolute emission reductions based on the local energy mix.
\end{abstract}


\section{Introduction}

Electric Vehicles (EVs) are meeting the expectation of significantly reducing greenhouse gas emissions, as their Global Warming Potential (GWP), expressed in $\mathrm{gCO_2e/vkm}$ (grams of carbon dioxide equivalent per vehicle-kilometre), is much lower than that of gasoline and diesel Internal Combustion Engine Vehicles (ICEVs) \cite{EC_2020}. For instance, the net lifecycle impact of new EVs in 2020 is lower than that of new gasoline and diesel cars in all European countries, except for Estonia. Based on EV lifecycle assessments, the most significant contribution to their GWP is caused by charging events \cite{EC_2020}, as energy generated from fossil fuels may be used for this purpose, and its carbon intensity is more significant than that of EV production, maintenance and end-of-life processes. Accordingly, the GWP of EVs is mainly a consequence of the carbon intensity of the power generation ($C_{gen}$) used for charging the vehicle (measured in $\mathrm{gCO_2e/kWh}$). 
While $C_{gen}$ is relatively low in European countries, the world average of $C_{gen}$ is almost double that of Europe \cite{EC_2020}, undermining the full potential of EVs in the sustainability transition. Power generation from renewable sources is increasing worldwide, and as a consequence, $C_{gen}$ is expected to decrease. In the meantime, however, while power generation leads to emissions, there is a need for strategies to minimize the emissions associated with charging electric vehicles.

In this paper, we demonstrate that i) leveraging simple domestic smart plugs to schedule charging at less carbon intensive times, can significantly reduce carbon emissions per unit of energy charged ($C_{EV}$ [$\mathrm{gCO_2e/kWh}$]), and ii) that the potential for emission reductions can be enhanced by optimising charging schedules multiple sessions ahead. Our results further highlight iii) the importance of virtuous user behaviour such as plugging in for longer time periods and reducing daily energy consumption, and that iv) regional differences in $C_{gen}$ over time affect the overall potential of smart charging for decarbonization.

\IEEEpubidadjcol
Unidirectional or ``smart charging'' is widely recognized as the most convenient way to accommodate the electrification of the mobility sector while mitigating the need for electricity network expansion \cite{Fraunhofer_2024}. Besides, smart charging may be also used to extend EV battery life by an average of 40\% through lowering the target state-of-charge (SOC) from 100\% to 80\% upon departure \cite{Fraunhofer_2024}. Roughly speaking, smart charging refers to the possibility to control the power that is used to charge a plugged vehicle, as an alternative to uncontrolled charging, where a vehicle is charged at maximum power from the moment it is connected, until it is fully charged. Smart charging has been investigated extensively in the literature for a plethora of applications, and most notably for load balancing \cite{Veldman2015, Chen2022}; peak shaving \cite{Leemput2014, PeakShaving_2015}; energy cost minimization \cite{Zhang_2018, Bashash_2014}; grid impact minimization \cite{Sastry_2024}; or for maximizing the integration of renewable energy \cite{Zhang_2014, Sponge}. A comprehensive review of smart charging strategies can be found in \cite{ITS_2024}, and, with a focus on distributed solutions, in \cite{Nimalsiri_2020}. Many references also explicitly focus on fleets of EVs, as more flexibility can be gained through collective control of a (large) number of vehicles \cite{Aygun_2024}.

In this paper we reconsider the unidirectional smart charging problem for a single vehicle, in a domestic overnight scenario, with the following main contributions with respect to the state of the art:
\begin{itemize}
\item In contrast to the common cost functions of interest, here, we minimize the net carbon emissions of the electricity used for charging. 
\item We solve the optimization problem for a sliding forecast horizon spanning multiple charging sessions (e.g., 4 days or one week), by applying Model Predictive Control (MPC) \cite{Maciejowski}. Particularly, we leverage the flexibility to schedule charging within charging sessions (intra-night flexibility) \emph{and} to postpone charging to subsequent sessions (inter-night flexibility), guaranteeing that each morning a minimum SOC is available. We show that the additional inter-night flexibility may lead to increased emission reductions.
\item Our results characterise the role of charging and driving patterns of EV users in effectiveness of the proposed smart charging strategy. This may inform the design of emission trading markets, reducing collective emissions by enabling the trading of carbon savings, motivating more flexible users to plug-in more and less flexible users to pay the price of emissions.
\item Using carbon intensity data from 14 different UK regions, we compare the emission reduction potential of the proposed method, highlighting significant variation between regions. 
\end{itemize}

This paper is organized as follows: in the next section, we formulate the carbon intensity minimization problem, and describe the MPC strategy for overnight charging control. In Section \ref{input_sequences}, we explain how to model the uncertain variables of the optimization problem, most notably future carbon intensity of energy, and driving patterns of EV users. In Section \ref{sec:results}, through numerical simulations we evaluate the effectiveness of the proposed charging strategy, and furthermore investigate the effect of different forecast horizons and varying charging flexibility on the potential for emission reduction. The results are concluded by highlighting regional differences in terms of emission reduction potential. Finally, in Section \ref{sec:conclusions} we discuss the main findings together with  possible future directions to expand and address the limitations of this work.

\section{The MPC-based charging strategy}
\label{sec:MPC}
\begin{figure}
    \centering
    \includegraphics[width=\linewidth]{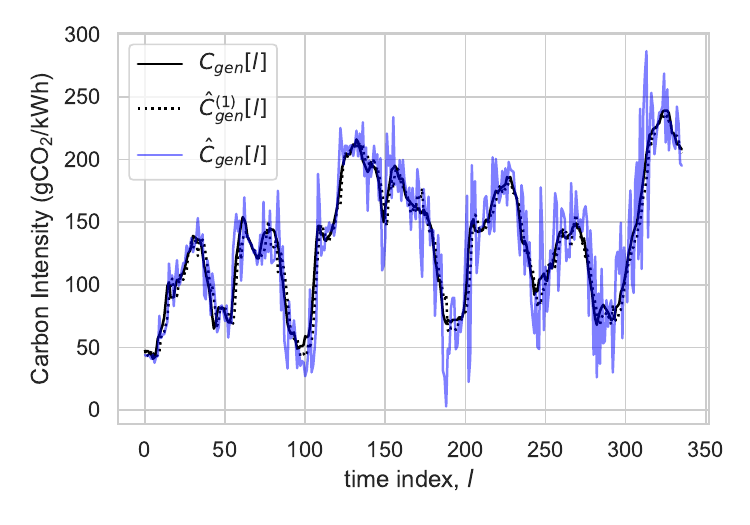}
    \caption{$C_{gen}[l]$, $l\in\{1,\ldots,  48 \times 7\}$ (black line): {\em actual\/} half-hourly carbon intensity measurements  for UK National Grid electricity generation during the first week  of January 2022, published by the National Energy System Operator (NESO) \cite{nationalenergysystemoperator};
    $\hat{C}_{gen}^{(1)}[l]$ (black dotted line): NESO's published sequence of {\em one-step-ahead\/} forecasts;   $\hat{C}_{gen}[l]$  (blue line): our  
    {\em $l$-step-ahead\/} forecast sequence (\ref{eq:forecastCgen}), estimated by transferring CarbonCast  side-information~\cite{maji2022}. 
    }
    \label{fig:C_gen_forecast_example}
\end{figure}

Figure \ref{fig:C_gen_forecast_example} shows the carbon intensity of electricity generation, $C_{gen}$ ($\mathrm{gCO_2/kWh}$), in the UK during the first week of January 2022. This sequence is made available half-hourly by the National Energy System Operator (NESO)~\cite{nationalenergysystemoperator}. Preliminary inspection reveals the characteristic daily pseudo-periodicity of this time series. In consequence, the range of $C_{gen}$, is large, reaching values that are approximately five times higher than the minima observed during the week. Carbon intensity depends on the electric load:  when consumption is greater, then more fossil-based power generation is required to supplement generation from renewable sources, and so carbon intensity increases. Correspondingly, $C_{gen}$ tends to be lower during the night, when the electric load is lower, and so generation from renewable sources can satisfy most of the demand for  electrical energy during these times. In addition to the daily pseudo-periodicity, we observe significant variation between days. 

The temporal variability of carbon intensity suggests that controlled EV charging---in which charging is scheduled during periods of low $C_{gen}$---has the potential to reduce the carbon intensity associated with EV usage \cite{li2023, mehlig2022}.

\subsection{The predicted net carbon objective function, $\hat{J}_0$}
We adopt a model predictive control (MPC) strategy \cite{Maciejowski} which schedules power delivery to the EVs during low-carbon  overnight charging intervals. We want to optimize this schedule in the context of the predicted variation in carbon intensity, $\hat{C}_{gen}$,  over multiple nights. Since a typical driver may not deplete the entire battery capacity during a day's driving, there is flexibility to delay charging to later (i.e.\ longer-horizon) nights, if $ \hat{C}_{gen}$ is  predicted to be high during the next (i.e.\ shorter-horizon) nights. Our predictive net carbon objective function should therefore meet the following criteria: 
\begin{itemize}
    \item minimize the net carbon emissions of charging the EV battery over a planning horizon---which we call the {\em forecast window\/}---of duration several nights;
    \item ensure that the SOC of the battery remains within safe operational limits at all times (assumed to be between 20\% and 80\% in this paper);
    \item ensure that the battery is sufficiently charged every morning to cover that day's predicted driving needs.
\end{itemize}
The MPC strategy then iteratively solves the minimization problem for net carbon over a sliding forecast window.

\subsubsection{Discretization of the forecast window}
Let the forecast window comprise $N \geq 1$ (24-hour) days, each divided into $c\equiv \frac{24}{\Delta_t}$ contiguous intervals of duration  $\Delta_t$ ( $\equiv 0.5$  hours in our implementation), and indexed by $l\in \{1, \ldots, c\times N\}$. The minimization of net carbon depends, of course, on $\hat{C}_{gen} [l]$, the predicted carbon intensity sequence for electricity generation during the forecast window, which is assumed to be constant during each $\Delta_t$ interval. Since these predictions are only required during the active (presumably overnight) charging sessions, it is convenient also to adopt a double-indexing convention,  
\begin{equation}
\hat{C}_{gen} [s,k] \equiv \left.\hat{C}_{gen} [l]\right|_{l\equiv c\times (s-1) +k}, \label{eq:estcarbon} 
\end{equation}
where $s\in\{1, \ldots, N\}$ and 
$k\in\{1, \ldots, c\}$; 
i.e.\ $\hat{C}_{gen}[l]$ can also be indexed  by the $k$ interval within the $s$th day of the forecast window. 

\subsubsection{Constrained minimization of net carbon}
We denote by $\hat{J}[l=0] \equiv \hat{J}_0$ the predicted net carbon ($\mathrm{gCO_2}$) associated with the multiple (typically overnight) charging sessions. Our objective is to minimize $\hat{J}_0$ by choosing an optimal sequence of charging powers, $\{P[s,k]\}$ (kW), during the $N$ charging sessions of the forecast window: 
\begin{align}
    \min_{\{P[s,k]\}}  & \hat{J}_0 \equiv \sum_{s=1}^{N} \sum_{k = \hat{k}_{s,b}}^{\hat{k}_{s,e}} \hat{C}_{gen}[s,k] \times P[s,k] \times \Delta_t. \label{eq:objective2}
\end{align}
On the $s$th 24-hour day (but typically during the night), the charging session will begin in the  $\hat{k}_{s,b}$th interval, and  end in the $\hat{k}_{s,e}$th interval, $\hat{k}_{s,e} > \hat{k}_{s,b}$, $ \forall s$.
$P[s,k]$ (kW) denotes the sequential charging powers during each $\Delta_t$-interval of each overnight charging session. These are again assumed to be constant over the intervals of duration, $\Delta_t$, and their optimization for minimal net carbon, $\hat{J}_0$, is the purpose of our scheme. 

The optimization (\ref{eq:objective2}) is subject to technology constraints associated with the domestic charging infrastructure and the EV battery itself. The SOC sequence of the EV battery (in \%) satisfies the following update equations:
\begin{eqnarray}
     SOC[s,k+1] &=& SOC[s,k] + \frac{P[s,k] \times \Delta_t \times 100}{B}, \nonumber\\
     && k\in\{\hat{k}_{s,b}, \ldots, \hat{k}_{s,e}-1\},\label{eq:soc_dynamics2} \\
     SOC[s+1,\hat{k}_{s+1,b}]  &=& SOC[s,\hat{k}_{s,e}] - \frac{\hat{E}_{s} \times 100}{B}, \nonumber \\
     && s\in\{1, \ldots, N-1\}. \label{eq:daily_consumption2}
    \end{eqnarray}
Here, $B$ denotes the battery capacity (kWh). (\ref{eq:soc_dynamics2}) models the increasing SOC sequence due to charging during the $s$th charging session, while (\ref{eq:daily_consumption2}) expresses the reduction in SOC between successive days, due to $\hat{E}_{s}$, being the estimated energy consumed from the battery because of the use of the EV during the $s$th day. 
Finally, we can express the constraints imposed on the optimization problem (\ref{eq:objective2}), both directly in terms of  $P[s,k]$, and indirectly in terms of  $SOC[s,k]$:
\begin{align}
& 0 \leq P[s,k] \leq \overline{P} \label{eq:power_bounds2}, \\
& \underline{SOC} \leq SOC[s,k] \leq \overline{SOC}, \label{eq:soc_bounds2}\\
& SOC[s,\hat{k}_{s,e}] \geq \underline{SOC}_{s}. \label{eq:morn_constr2} 
\end{align}
In  (\ref{eq:power_bounds2}), $\overline{P}$ is the nominal maximum power  rating (kW) of the domestic infrastructure.
Meanwhile, $(\underline{SOC},\overline{SOC}) \equiv (20,80)\%$ (\ref{eq:soc_bounds2}) is the interval within which the SOC should be maintained so as to maximize the lifetime of the battery. Finally, 
$\underline{SOC}_{s}$ in \eqref{eq:morn_constr2} is the minimum SOC required at the end of the $s$th overnight charging session in order to meet the  energy consumption, $\hat{E}_{s}$ (kWh) (\ref{eq:daily_consumption2}), of the day ahead. 

The net carbon objective, $\hat{J}_0$ (\ref{eq:objective2}), depends, not only on the predicted carbon intensity sequence, $\hat{C}_{gen}[l]$, during the charging sessions, i.e.\ $\hat{C}_{gen}[s,k]$ (\ref{eq:estcarbon}), but also on specification of the start and end times of each of these sessions, via $\hat{k}_{s,b}$ and $\hat{k}_{s,e}$, respectively. It also depends  on the estimated energy consumption sequence, $\hat{E}_s$, via (\ref{eq:daily_consumption2}).  
Our technique for computing $\hat{C}_{gen}[l]$ from an external database will be  explained in Section~\ref{sec:carbon_intensity_forecast}. Our computation of   $\hat{k}_{s,b}$ and $\hat{k}_{s,e}$ will be explained in  Section~\ref{Uncertain_Time}, and of  $\hat{E}_{s}$ in Section~\ref{sec:Esequence}.

\section{Estimation of the Input Sequences}
\label{input_sequences}

\subsection{The long-range carbon intensity forecast, $\hat{C}_{gen}[l]$}
\label{sec:carbon_intensity_forecast}
The objective \eqref{eq:objective2} minimizes the predicted net carbon, $\hat{J}_0$, at arbitrary datum, $l=0$, over the subsequent $N$-day forecast window, in which the predicted carbon intensity is $\hat{C}_{gen}[l]$, $l\in\{1, \ldots, c\times N\}$ (\ref{eq:estcarbon}). The minimization is repeated every $\Delta_t$ hours, by sliding the forecast window forward in time, i.e.\  $l \leftarrow l+1$.    
Note that NESO makes available the {\em actual\/} half-hourly carbon intensity sequence, $C_{gen}[l]$, as well as its half-hourly, short-range forecast  up to $N=2$ days ahead. However,  only the one-step-ahead NESO prediction is actually stored. We will  specifically denote this NESO-stored one-step-ahead prediction of carbon intensity at time interval, $l$, by $\hat{C}_{gen}^{(1)}[l]$ (Figure~\ref{fig:C_gen_forecast_example}). 
Therefore, our requirement for the carbon minimization objective (\ref{eq:objective2}) is to synthesize the $l$-step-ahead  carbon intensity forecast sequence, $\hat{C}_{gen}[l]$, over the long-range  forecast window,  in a manner which captures its increasing variance with $l$, as seen in the example in Figure~\ref{fig:C_gen_forecast_example}.

To summarize,  NESO's publicly available data~\cite{nationalenergysystemoperator} provides the following sequences at any arbitrary datum ($l=0$):
\begin{itemize}
    \item $C_{gen}[l]$, the actual half-hourly (average) carbon intensity sequence for the UK national grid, indexed by $l\in\{1, \ldots, c\times N\}$;
    \item $\hat{C}_{gen}^{(1)}[l]$, the associated {\em one-step-ahead\/} forecast of the carbon intensity in time interval, $l$.  

\end{itemize}

\subsubsection{Relative error of the $l$-step-ahead forecast}
This is defined as follows:
\begin{equation}
    \hat{\epsilon}[l] = \frac{\hat{C}_{gen}[l] - C_{gen}[l]}{C_{gen}[l]},
    \label{eq:forecast_errors}
\end{equation}

Hence:
\begin{equation}
    \hat{C}_{gen}[l] = C_{gen}[l]\left(1+\hat{\epsilon}[l]\right).
    \label{eq:forecast_sequence}
\end{equation}
Hence, if we can estimate---from available data---the $l$-step-ahead forecasting error,  $\hat{\epsilon}[l]$, we can retrospectively synthesize the associated multi-step-ahead (i.e.\ long-range) noisy forecast---required in our objective (\ref{eq:objective2})---from the NESO-stored sequence  of actual carbon intensities,  $C_{gen}[l]$.

\subsubsection{Transferring CarbonCast~\cite{maji2022} side-information for linear scaling of the forecast error }
The mean absolute percentage error (MAPE) statistics for carbon intensity forecasts up to four days have been published in~\cite{maji2022} (Table~6), using those authors' own CarbonCast forecasting technique along with two alternative forecasting techniques. As explained below, these statistics  support a hypothesis of linear growth in the expected MAPE of these long-range forecasts. Under this hypothesis, we therefore write
\begin{equation}
    \lvert\hat{\epsilon}[l]\rvert = \lvert\hat{\epsilon}[1] \rvert\times\left(1+\hat{\lambda}\times (l-1)\right),
    \label{eq:error_magnitude}
\end{equation}
where $\hat{\lambda}$ is the estimated constant rate of  growth of the forecast MAPE. 

Recall that the one-step-ahead relative prediction error, $\hat{\epsilon}[1]$, can be computed directly from the published and stored NESO sequences, $C_{gen}[1]$ and $\hat{C}_{gen}^{(1)}[1]$, via (\ref{eq:forecast_errors}). 

Obviously, the absolute error sequence (\ref{eq:error_magnitude}) suppresses the sign (i.e.\ polarity) of each $\hat{\epsilon}[l]$,  which we denote by $a[l] \in\{+1, -1\}$. We recover these in our simulated forecasts via iid Bernoulli trials of probability $\frac{1}{2}$. Then:

\begin{equation}
    \hat{\epsilon}[l] = a[l]\times \lvert\hat{\epsilon}[l]\rvert.
    \label{eq:error_sign}
\end{equation}
Inserting (\ref{eq:error_sign}) and (\ref{eq:error_magnitude}) into (\ref{eq:forecast_sequence}), we obtain a formula for synthesizing the long-range carbon intensity forecast sequence in terms of the published NESO data, with knowledge transfer (into $\hat{\lambda}$) from the CarbonCast data~\cite{maji2022}: 

\begin{equation}
    \hat{C}_{gen}[l] = C_{gen}[l] \times \left[1+a[l]\times \lvert\hat{\epsilon}[1]\rvert \times \left(1+\hat{\lambda} \times (l-1)\right)\right].
         \label{eq:forecastCgen}
\end{equation}

\subsubsection{Computation of $\hat{\lambda}$}
We compute the  error growth rate per half-hour interval to be  $\hat{\lambda} \approx 9.97\times10^{-3}$, based on the published forecasts in \cite{maji2022} (Table~6), in which the MAPEs exhibit  an approximately linear (albeit slightly concave) increase up to a horizon of four days,  and across multiple regions (excluding the UK\footnote{UK-specific data are not reported, and so we estimate them via  the average growth rate across the actually reported regions. Calibration with UK-specific data may refine this approach in the future.}). The linear model adopted here is a reasonable simplification which leads to a conservative estimate, since it overestimates long-range forecast errors. Figure \ref{fig:C_gen_forecast_example} shows a typical comparison between our synthetic forecast, $\hat{C}_{gen}[l]$ (\ref{eq:forecastCgen}), and the {\em  actual\/} stored NESO carbon intensity sequence, $C_{gen}[l]$.

As $l$ increases, errors accumulate in the synthetic forecast, approximating the growing uncertainty over longer forecast horizons. Overall, the method described here provides a practical way to evaluate our charging strategy subject to the increasing forecast error exhibited by the  published data.\newline
\textbf{Remark:} We note that the value of the error growth rate per half-hour may significantly change from one area to another area. In particular, as we have already mentioned, the carbon intensity largely depends on how much energy has been generated from renewable sources, and it may be simpler to predict when energy is mainly generated from solar plants, for instance, while it may be harder to predict when it is mainly generated from wind farms, where generation forecasts are less accurate. Here, we rely on the estimate of $\hat{\lambda}$ from \cite{maji2022}, obtained as an average value amongst six different geographic regions across US and Europe.

\subsection{Expected  start and end times, $\hat{k}_{s,b}$ and $\hat{k}_{s,e}$, of charging sessions }
\label{Uncertain_Time}
The start and end times of the charging sessions in (\ref{eq:objective2})---quantized and indexed by $l=\hat{k}_{s,b}$ and $l=\hat{k}_{s,e}$ for the $s$th session, $s\in\{1, \ldots, N\}$---are chosen in order to represent a typical daily commuting pattern, and are based on average values from a data set of domestic charging events \cite{departmentfortransport2018}. These times are modelled via normal distributions with a one-hour standard deviation, centred around 18:00 (start) and 09:00 (end). We then choose the $s$-invariant start time ($\hat{k}_{s,b}\equiv \hat{k}_{b}$) and end time ($\hat{k}_{s,e}\equiv \hat{k}_{e}$)  as the 98th percentile (i.e. 20:03)  and 2nd percentile (i.e.\ 06:57), respectively.   
These choices lead to  a conservative design (i.e. over-design) of the charging power sequence, $P[s,k]$, in (\ref{eq:objective2}), ensuring that    the EV is  sufficiently charged with high probability, even if it is plugged in unusually late, and plugged out unusually early.

\subsection{Expected Daily Energy Consumption, $\hat{E}_s$}
\label{sec:Esequence}
The daily estimated energy consumption values, $\hat{E}_s$, for the $N$ charging sessions (\ref{eq:daily_consumption2}),  are also  computed  $s$-invariantly via a stationary normal model:
\[
E_s \stackrel{iid}{\sim} \mathcal{N}(5.8,2.67^2).
\]
Here, the mean and standard deviation are informed by the  National Travel Survey (NTS) dataset of the Department for Transport~\cite{departmentfortransport2018}. The actual $s$-invariant value, $\hat{E}_s \equiv \hat{E}$, inserted into (\ref{eq:daily_consumption2}), is again chosen conservatively, as the 98th percentile value ($\hat{E}\approx 11.28$ kWh), to ensure the over-design of the charging power sequence, $P[s,k]$, in (\ref{eq:objective2}).
We also assume that outlier high-consumption events are known in advance (e.g.\ the day before), and so the driver can override the default assumption, above, by increasing the minimum SOC requirement---$\underline{SOC}_s$ (\ref{eq:morn_constr2})--- for that day.

\section{Simulated Charging Scenarios}
\label{sec:results}
We now evaluate the effectiveness of the proposed MPC strategy for minimizing carbon emissions in the charging processes(section \ref{sec:MPC}), through a number of different simulations. 

All case studies are simulated for a period of one year based on the carbon intensity of the UK national electricity grid. Vehicle and charge point parameters are specified to resemble typical mid-range EVs and domestic charging stations. The battery capacity is set at 50kWh and the maximum charging power at 10kW. 

\subsection{MPC Optimization}
\begin{figure}[h]
    \centering
    \includegraphics[width=\linewidth]{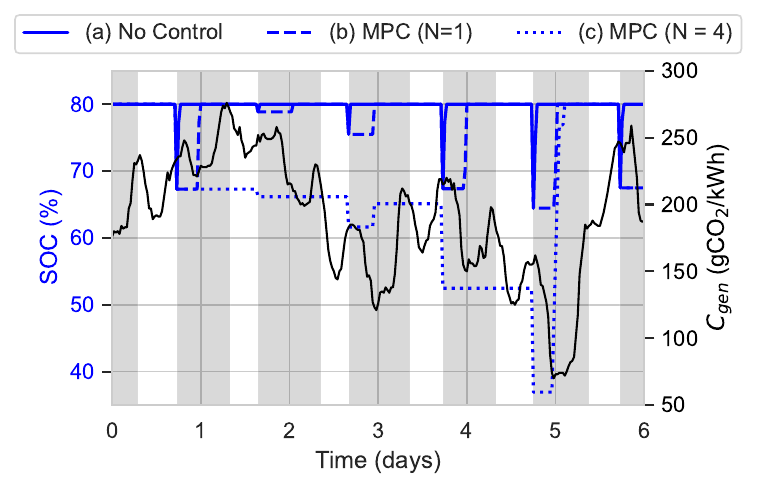}
    \caption{State of Charge (SOC) profiles over six simulation days for three charging strategies: uncontrolled (solid blue), MPC with a 1-day horizon (dashed blue), and MPC with a 4-day horizon (dotted blue). The black line shows the carbon intensity signal (C$_{\text{gen}}$). Gray background indicates night charging windows, during which the vehicle is plugged in; The figure highlights how predictive strategies shift charging to periods of lower carbon intensity.}
    \label{fig:SOC_over_time}
\end{figure}

\begin{figure}[h]
    \centering
    \includegraphics[width=\linewidth]{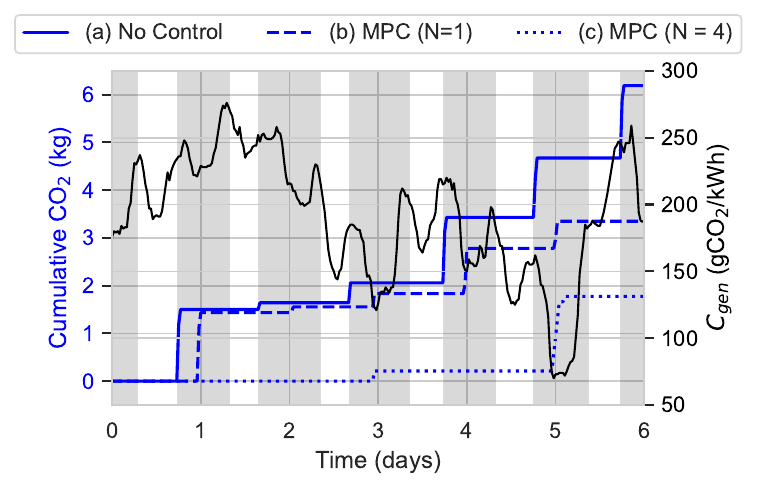}
    \caption{Cumulative CO$_2$ emissions over six simulation days for the same three charging strategies. Emissions increase stepwise with each charging event. MPC strategies result in significantly lower cumulative emissions.}
    \label{fig:CO2_over_time}
\end{figure}

In the first simulation, we compare (a) an uncontrolled strategy, where a vehicle immediately starts the charging process up to $\overline{SOC}$ as soon as it is connected at home, (b) a single session predictive control approach, i.e., the strategy outlined in  section \ref{sec:MPC} with $N = 1$, and (c) the same MPC strategy with $N = 4$.
Figures \ref{fig:SOC_over_time} and \ref{fig:CO2_over_time} explain the differences among the three strategies for a single week of simulation, for clarity. In the figures, the gray background rectangles identify night charging sessions, while the white background relates to daily hours, when the vehicles can not be recharged since it is not connected. The black curve depicts the actual carbon intensity during the week.

Figure \ref{fig:SOC_over_time} depicts the evolution of the SOC (in blue), as a function of the C$_{gen}$ signal, in black. As can be seen the uncontrolled strategy (solid blue line) immediately charges the vehicles at the beginning of the gray rectangles (night charging sessions), while the MPC strategy (dashed line) shifts the charging process to the most convenient moment of the night, when the carbon intensity is at the lowest value. The MPC strategy with a longer future horizon of time (N = 4) prefers not to charge the vehicle during some of the nights, if the carbon intensity is not convenient, and waits until the most convenient night for charging. In all cases, the morning SOC (i.e., at the beginning of the white rectangles) is always above 50\%, which here had been selected as the minimum desired SOC of the owner every morning.

Figure \ref{fig:CO2_over_time} depicts the cumulative carbon intensity of the three strategies during the same one week period, and it increases to a new value every time a charging event is concluded. The MPC strategies obtain significantly reduced amounts of $CO_2$, and with a forecast horizon of N = 4 days, the amount is about $50\%$ of the amount of the uncontrolled case. 

Table \ref{tab:results_table} summarizes the outcomes of the simulation for a whole year, adding the further cases of N = 2 and N = 7. As can be noticed, it is confirmed that after one year the amount of $CO_2$ is halved when a future horizon of 4 days is selected. Interestingly, increasing the future horizon of time to 7 days does not seem to provide further improvements, but instead may lead to higher carbon intensity, since the additional flexibility of the longer horizon is counterbalanced by the increased inaccuracy of the estimate of the carbon intensity. 

\

\begin{table}
    \centering
\caption{Aggregate simulation results for four different smart charging strategies (2022, UK National Grid)}
\label{tab:results_table}
    \begin{tabular}{llll}
\hline
         Strategy

&  $C_{EV}$ ($gCO_2$/kWh)
& \% Reduction\\
\hline
         Uncontrolled
&  195.74
& 0.0
\\
         MPC (N = 1)
&  141.96
& 27.48
\\
         MPC (N = 2)
&  116.69
& 40.38
\\
         MPC (N = 4)
&  104.76
& 46.48
\\
MPC (N = 7)
&  113.50
& 42.01\\
\hline
    \end{tabular}
    \end{table}

\subsection{Rewarding flexibility in emission trading markets}
\label{sec:flexibility}
\begin{figure}
    \centering
    \includegraphics[width=\linewidth]{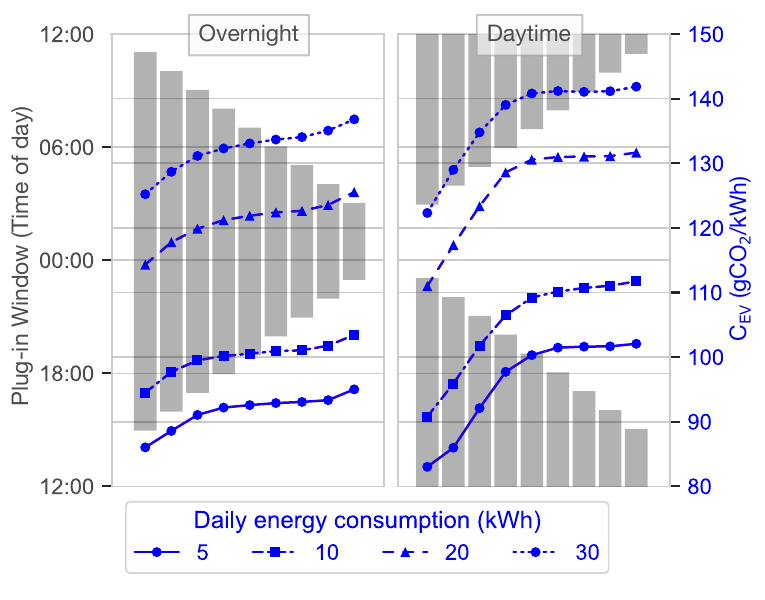}
    \caption{Impact of user flexibility on average carbon intensity ($C_{EV}$) during one month of simulated EV charging (January 2023, UK national carbon intensity data). Each gray bar represents a different daily plug-in time window, ranging from 20 to 4 hours. Four lines correspond to daily energy demands of 5, 10, 20, and 30 kWh. Shorter plug-in durations and higher energy demands both increase average carbon intensity. Daytime charging scenarios (right side) typically result in significantly higher emissions than overnight charging at comparable plug-in durations.}
    \label{fig:user_flexibility_results}
\end{figure}

The carbon intensity that is achieved when employing such smart charging approaches ultimately depends on the driving and charging patterns of the user. For example, if the vehicle is plugged in for a long duration, or at the right times, the controller is more likely able to charge during periods of low carbon intensity. On the other hand, if the vehicle is connected infrequently, such low-carbon periods may be missed, resulting in a higher average carbon intensity. Similarly, if the user consistently uses little energy for driving, then the smart charging algorithm has a greater flexibility to delay charging until convenient periods of low carbon intensity are met. On the other hand, if the battery is depleted every day, then there is no such flexibility, and the battery may have to be charged for the whole night.
Figure \ref{fig:user_flexibility_results} illustrates how the average carbon intensity of energy charged by an EV using multi-session predictive control (N=4) varies, depending on both the plug-in time frame, as well as the daily energy demand of the driver. The results are based on one simulation month (January 2023) with UK National grid carbon intensity. The grey bars illustrate the daily plug-in periods, with plug-in durations ranging between 20 and 4 hours, decreasing in steps of two hours. Overnight charging scenarios are represented on the left side of the figure, and daytime charging scenarios on the right side. The four lines show the average carbon intensity achieved based on daily energy demands of 5,10,20, and 30 kWh---approximately equivalent to a daily driving distance of 25, 50, 100, and 150 km. We can see, that plugging in for shorter periods generally leads to higher carbon intensity for all energy demands---slightly higher ($\approx 8\%$) for overnight scenarios and significantly higher ($\approx 25\%$) for daytime charging scenarios. With the exception of the first two daytime scenarios (20 and 18 hours total plug-in duration), charging during the day leads to higher carbon intensity than charging over night, even if plug-in times are longer. The figure further shows that EVs with small daily energy demands (e.g. 5kWh) can charge less carbon intensive energy on average. In comparison, the energy charged by drivers that require 30kWh of energy every day is on average $27.3 \%$ more carbon intensive.

Overall, these findings highlight that individual driving and charging patterns influence the carbon intensity of EV charging. A natural next step is thus to explore how to incentivize ``virtuous" behaviours, i.e., little usage of the vehicle or leaving it plugged in as much as possible, that may lead to reduced charging-related emissions. One way to do this could be through emission markets or pooling mechanisms, which create financial incentives for reducing EV-related emissions below a threshold. A similar scheme is well-established in the EU to reduce CO$_2$ emissions from passenger cars and light commercial vehicles. Regulation 2019/631 sets decreasing fleet-wide emission targets (e.g., 95 g CO$_2$/km for 2020-24) for vehicles sold by car manufacturers in a given year\cite{europeanunion2019}. To achieve these targets and avoid fines, manufacturers can form pools, allowing them to jointly meet emission targets. Within pooling arrangements, manufacturers with higher emissions purchase emission credits from manufacturers exceeding the target (i.e. Tesla, Polestar, Volvo \cite{reuters2025}). Several manufacturers are reported to utilize such pooling arrangements to comply with EU regulation. For instance, Stellantis, Toyota, Ford, Mazda, and Subaru plan to pool their fleets with Tesla; Similarly, Mercedes has entered a pooling agreement with Polestar, Volvo Cars, and Smart \cite{reuters2025}. 
Similar schemes could be adopted for charging EVs as well, with the goal to incentivize EV users to plug in overnight and for longer periods, allowing them to benefit financially from selling emission reduction credits to users with less flexibility.

\subsection{Regional Variation}
\begin{figure}
    \centering
    \includegraphics[width=\linewidth]{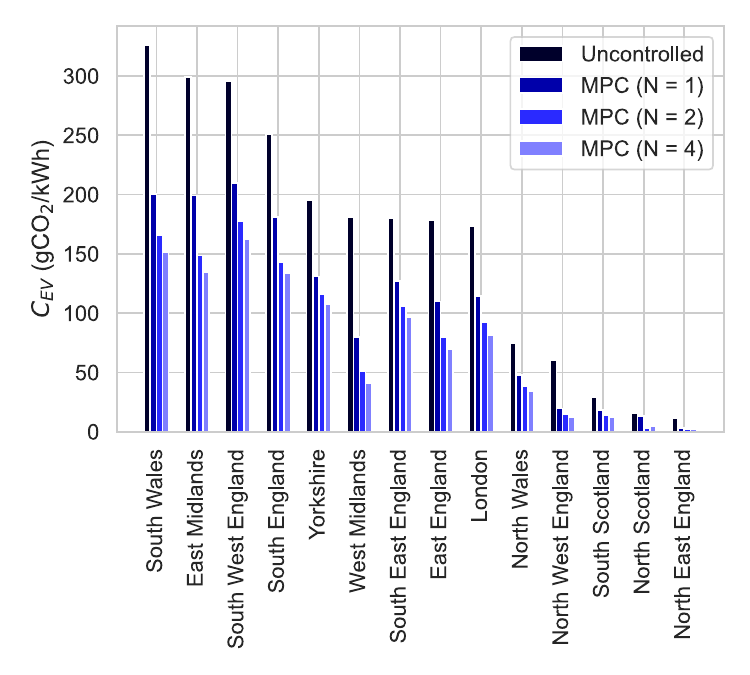}
    \caption{Regional variation in average carbon intensity ($C_{EV}$) of EV charging across 14 UK regions during January 2023. Results are shown for uncontrolled charging and three predictive MPC strategies with forecast horizons $N = 1, 2, 4$ days. Smart charging reduces emissions in all regions, with relative reductions ranging from 16\% to 83.9\%. Longer prediction horizons yield larger reductions across regions.}
    \label{fig:regional_results}
\end{figure}

Regional factors also shape the effectiveness of smart charging. As highlighted in the introduction, the GWP of power generation varies significantly between countries and regions. In this section, we simulate the potential of MPC-based smart charging with prediction horizons $N=1,2,4$ days in 14 different UK regions. UK Regional data on the GWP of electricity in 2023 for this simulation is published by the National Energy System Operator in half-hourly timesteps \cite{nationalenergysystemoperator}.
Notably, the GWP of electricity shows significant regional variation and is much lower in northern regions, such as Scotland, North England and North Wales. On average, each unit of electricity in South Wales emits more than three times more carbon emissions when compared to these northern regions. Fig.\ref{fig:regional_results} shows the average carbon intensity achieved by uncontrolled charging versus MPC charging (Section \ref{sec:MPC}) during a simulation of one month (January 2023).
It is evident, that (1) Smart charging reduces the carbon intensity of EV charging effectively across all regions of the UK, with relative improvements ranging between 16 and 83.9\% depending on region and prediction horizon. (2) Longer prediction horizons (e.g., N=4) generally lead to higher reductions in GWP, with the average percentage reduction for prediction horizons of N=1,2,4 being 39.2\%, 55.5\%, and 59.2\% respectively. (3) The absolute reduction in GWP achieved through smart charging is particularly relevant in those regions with a higher average GWP of electricity. For example, the absolute reduction in average GWP of energy charged in South Wales using Smart Charging with 4 day prediction horizon is 175 gCO$_2$/kWh, while the respective absolute reduction in North East England is only 9.8 gCO$_2$/kWh. Here, even uncontrolled charging is relatively clean at 11.7 gCO$_2$/kWh, leaving little potential for improvement. Finally, (4) the relative reduction achieved through smart charging against uncontrolled charging varies between regions, suggesting that the individual carbon intensity profile of a region determines the relative emission reduction potential for smart charging. However, in all cases, independently from the specific carbon intensity of power generation, the MPC scheme appears to significantly reduce the emissions of the uncontrolled strategy.

\section{Conclusions}
\label{sec:conclusions}
This paper presented a multi-session MPC strategy for overnight EV charging that leverages forecasts of grid carbon intensity to reduce charging-related CO$_2$ emissions. Results show that scheduling charges over multiple sessions (up to one week ahead) outperforms session-by-session MPC and uncontrolled ``plug-and-charge" approaches. These savings are enabled by scheduling charging power during multi-day minima of grid carbon intensity and by exploiting the flexibility to delay charging of energy that is not required for next day's driving.

Our simulations confirm, that carbon emission reductions depend on the energy need and charging behaviour of the user, as well as the GWP profile of the local power grid. Longer and well-timed plug-in durations as well as modest daily energy requirements contribute to enhanced emission reductions, and particularly relevant absolute savings can be achieved in regions with high average GWP of electricity.

Our work suggests that it is very simple to achieve significant improvements of the carbon intensity of EVs, however a few aspects need to be better investigated. First of all, a massive adoption of the proposed strategies may indeed have an impact on the $C_{gen}$ curve (i.e., if a huge number of EVs synchronously choose the same moment when to charge the vehicle, then this may cause an increase on the expected $C_{gen}$ signal). In this case, when the multi-session optimization approach is scaled to fleets of EVs, it may be convenient to design an aggregator entity to handle the charging process, and possibly to supervise the emission trades between EV participants.

\bibliographystyle{ieeetr}
\bibliography{MPC_Charging}

\vfill

\end{document}